\documentclass{ws-p8-50x6-00}
\makeatletter
\newcommand\preprintnote{%
        To appear in the Proceedings of 9th International Workshop on Deep
        Inelastic Scattering and QCD (DIS 2001), Bologna, Italy, 27 Apr -- 1
        May, 2001
}
\newcommand\preprint{
   \def\ps@plain{%
       \let\@mkboth\@gobbletwo
       \let\@oddhead\@empty
       \def\@oddfoot{%
           \lower1pc\hbox{\fbox{%
              \parbox{\hsize}{{\it \preprintnote}}%
           }}
       }%
   }
   \def\ps@simple{%
       \let\@mkboth\@gobbletwo
       \let\@oddhead\@empty
       \def\@oddfoot{\hfill {\bf \thepage}\hfill}%
   }
   \AtBeginDocument{
       \pagestyle{simple}
   }
}
\makeatother
\preprint

\newcommand{\mbf}[1]{\mbox{\boldmath $#1$}}
\begin{document}

\title{Odderon in QCD}

\author{G. P. Vacca}

\address{ II. Institut f\"ur Theoretische Physik,
Universit\"at Hamburg, Luruper Chaussee 149, D-22761 Hamburg, Germany\\ 
E-mail: vacca@bo.infn.it}

\maketitle

\abstracts{A review on how the Odderon idea does appear in QCD is given. 
In the last years it has been developed a non-perturbative QCD approach 
based on the stochastic vacuum model and a perturbative one based 
on resummation techniques in the small $x$ QCD region. 
Last developments on the perturbative analysis are shown in some details,
in particular in application to diffractive $\eta_c$ production. }

\section{Introduction}
It is remarkable that QCD predicts the existence of the Odderon.
The idea of the Odderon \cite{Lukaszuk:1973nt}, the partner of the Pomeron which
is odd under parity $P$ and charge conjugation $C$ (like the photon),
is related to the possibility that the real part of a scattering amplitude increases with
energy as fast as the imaginary part.
The scattering amplitude in the complex angular momentum plane possesses a
rightmost singularity (pole) near $j=1$. In the even (under crossing) amplitude
such a singularity is associated to the Pomeron and gives a mostly imaginary contribution,
while in the odd case one has a mostly real contribution which is associated to the
Odderon. The position of the singularity is also called intercept and it is related
to the asymptotic behaviour of the cross section.

The very successful theory for the strong interaction, QCD,  predicts
the Pomeron existence, in the most simple version as a two gluon exchange in colour
singlet state, thanks to the vector nature of the gluon.

The fact that the internal gauge symmetry group of QCD has rank greater than one
permits to construct from three gluons a $C$-odd state which can be associated to
the Odderon.

One can see this fact considering the $SU(3)_C$ gauge group associated
to the gluon field $A_\mu=\sum_a A_\mu^a t_a$.
Since under charge conjugation one has $A_\mu \to -A_\mu^T$, the two possible
independent invariants, constructed by three gluon fields, are
$Tr( [A_1,A_2]A_3 )$ and $Tr( \{A_1,A_2\}A_3 )$, which are respectively even and odd
under charge conjugation.
Therefore the Odderon will be related to the composite operator
$O_{\al\beta\gamma}=d_{abc}A_\al^a A_\beta^b A_\gamma^c$.

What is challenging the physics community is that experimentally there is
till now no clear evidence of the Odderon. Among the ways to look at his presence
there is the comparison of the total and/or elastic cross sections for direct
and cross-symmetric scattering processes, like for example in the case of $pp$ and
$p\bar{p}$ scattering. It was infact in this context that the Odderon was originally
introduced.

Another class of scattering processes, where the Odderon contributes, is when one or two
of the incoming scattering particles, of definite C-parity, goes into a state of
opposite C-parity under scattering.
One requires a rapidity gap which allows to separate the outgoing scattering states.
A typical example is given by the reaction
\bea
\gamma \, (\gamma^*) + p \to PS \, (T) + p \, (X_p), \nonumber
\eea  
where a photon scatters a proton and a pseudoscalar or a tensor meson is produced in
the photon fragmentation region, well separated in rapidity from the proton or its debris
($X_p$). This process has started to be analyzed at HERA \cite{Schafer:1992pq}.
The $\gamma \gamma$ scattering process \cite{Ginzburg:1992mi} could be also
interesting, even if the cross section involved are much smaller .

In this direction perturbative analysis have been performed in the study of
$\eta_c$ production in DIS with an Odderon made by three simply uncorrelated gluons
\cite{Czyzewski:1997bv,Engel:1998cg}
and later by considering the resummed QCD interaction in LLA \cite{Bartels:2001hw}.
Some details of the last approach are given in the section 4.

Non perturbative studies have been carried on for the production of light mesons
($\pi^0$, $f_2$) \cite{Heidelberg}.
This general non perturbative approach will be sketched in section 3.
The $\pi^0$ production process has been very recently analyzed at HERA
by the H1 collaboration, the Odderon has not been seen and it has been put an upperbound
on the cross section ten times
smaller than the predicted cross section \cite{odderonH1}, setting a new challenge for the
theory understanding.

Another interesting proposal, based on a more phenomenological approach, has been the
study of charge asymmetry in charm states due to Pomeron-Odderon interference
\cite{Brodsky:1999mz}.

\section{Perturbative QCD Odderon in LLA}
A scattering process dominated by the Odderon exchange can be described in the high energy
limit, in the context of $k_T$ factorization, by an amplitude
\bea
A(s,t) = \frac{s}{32}\frac{1}{16}\frac{N_c^2-4}{N_c}\frac{N_c^2-1}{3!}\frac{1}{(2\pi)^8}
\langle \Phi^i_{\gamma}|G_3|\Phi_p\rangle.
\label{ampli}
\eea
At lowest order, provided the strong coupling $\alpha_s$ is small,
one has a simple three uncorrelated gluon exchange, i.e. the Green function $G_3$, which is
convoluted with the impact factors, is constructed, simply with 3 gluon propagators.
Therefore, in momentum representation
$G_3^{(LO)}=\delta^{(2)}(\mbf{k}_1-\mbf{k}'_1)\delta^{(2)}(\mbf{k}_2-\mbf{k}'_2)
1/\mbf{k}_1^2\mbf{k}_2^2\mbf{k}_3^2$.

In the high energy limit, when all other physical invariants are much smaller,
a LLA resummation of the contributions of the order $(\alpha_s \ln s)^n$,
which is not small, can be performed and one obtains,through $G_3$,
an effective evolution in rapidity.
The same resummation for the two gluon exchange has lead to the BFKL \cite{BFKL}
equation where it appears the kernel of the integral equation for the 2-gluon Green function
that, in the colour singlet state, describes the perturbative QCD Pomeron in LLA.
The same equation in the colour octet state has a simple eigenstate which corresponds
to the reggeized gluon, which is in general at high energies a composed object.
This fact is seen as a self consistency requirement and it is called bootstrap.
In NLA \cite{Fadin:1998py}, where one is resumming also the contribution of order
$\alpha_s^n (\ln s)^{n-1}$, all the same concepts, reggeization included \cite{bootstrap},
apply.

The general kernel for the $n$-gluon integral equation for the Green function in LLA is
given by the BKP equation \cite{BKP}. In the large $N_c$ limit and for finite $N_c$ when
$n=3$, it possesses remarkable symmetry properties:
discrete cyclic symmetry, holomorphic separability, conformal invariance, integrability,
duality \cite{symmetries} and also a relation between solutions with different $n$
exists \cite{Vacca:2000bk}, which is a direct consequence of the gluon reggeization.

The Odderon states in LLA must be symmetric eigenstates of the operator
$K_3=1/2(K_{12}+K_{23}+K_{31})$ constructed with the BFKL kernel $K_{ij}$ for two
reggeized gluons in a singlet state.
Using the conformal invariance and integrability properties a set of eigenstates
has been found \cite{Janik:1999xj}, which have a maximal intercept below one. 

Using the gluon reggeization property (bootstrap) a new set of solutions was later found  
\cite{Bartels:2000yt}, characterized by intercept up to one, therefore dominant at high
energies. Moreover for the particular impact factor which couples a photon and an $\eta_c$ to
the Odderon the LLA calculation has shown that this second set of solution is relevant while
the previous one decouples.
We present here these Odderon states, since they will be used in section 4. 
In momentum representation they are given by $E_3^{(\nu,n)}$ such that
\bea
\mbf{k}_1^2 \mbf{k}_2^2 \mbf{k}_3^2 \, E_3^{(\nu,n)}(\mbf{k}_1,\mbf{k}_2,\mbf{k}_3)=c(\nu,n)
\sum_{(123)} (\mbf{k}_1+\mbf{k}_2)^2 \mbf{k}_3^2 \, 
E^{(\nu,n)}(\mbf{k}_1+\mbf{k}_2,\mbf{k}_3),
\label{oddwave}
\eea
where $c(n,\nu)$ is a normalization factor, $E$ is a BFKL pomeron eigenstate
and the conformal spin $n$ is odd.
The full Green function is constructed summing over all such states but in the high
energy limit the asymptotic behaviour can be studied for conformal spin $n=\pm 1$
and performing the saddle point integration around $\nu=0$.
 
\section{Non perturbative QCD Odderon}
A very briefly sketch on the non perturbative QCD framework used for Odderon studies
\cite{Heidelberg} is given in the following.

A first ingredient is the choice of the eikonal semiclassical
approximation\cite{Nachtmann:1991ua} for high energy scattering of quarks, while, at first, 
a full quantum colour field behaviour is considered.
In particular each quark which scatters on a colour field picks up a non abelian eikonal phase
$V=P \exp{[-i g \int_\Gamma d z^\mu \mbf{A}_\mu(z)]}$. 

The functional integral on the physical gluon field is estimated using the
stochastic vacuum model (SVM) \cite{Dosch:1988ha}, i.e.
the calculation of any correlation functions of gluon field strength is
associated to a gaussian stochastic process with finite correlation length
and, therefore, expanded as
$\langle F \cdot F \cdots F \rangle = \sum \, \prod \langle F \cdot F \rangle$. 
After some other assumptions and relating the basic two point function
$\langle 0| F \cdot F|0 \rangle$ to the gluon vacuum condensate, a dipole - dipole or
dipole-tripole (as Wegner-Wilson loops) scattering amplitude at fixed transverse size can
be computed expanding the ordered exponential.
Mesons (barions) are described in term of dipoles (tripoles) and transverse wave functions
\cite{Dosch:1994ym}.

When expanding the exponentials in the eikonal phases, terms of the kind
$\langle Tr(F \cdot F) Tr(F \cdot F)\rangle$
give imaginary contribution and are associated to the Pomeron.
Instead the real Odderon contribution is given by subsequent terms of the kind
$\langle Tr(F \cdot F \cdot F) Tr(F \cdot F \cdot F)\rangle$, in particular by the piece
with the $d_{abc}d_{abc}$ colour structure. 

In this approach the energy dependence can be introduced in a
phenomenological way. In general a diquark structure of the hadrons is preferred.
Regarding Odderon driven processes, in particular the production of light mesons
in DIS\cite{Heidelberg} has been studied($\pi^0$, $f_2$ with $N^*$ resonances production).
Predictions at HERA energies are
$\sigma^{\cal O}_{\gamma p \to \pi^0 N} \approx 400$ nb and
$\sigma^{\cal O}_{\gamma p \to f_2 N} \approx 21$ nb.

The first process has been recently analyzed at HERA by the H1 collaboration and there is now
an upperbound on the cross section of around $39$ nb\cite{odderonH1}.
It is important to understand such a big discrepancy. One possible source of error
comes from the parameter fixing in SVM,
but the most serious one seems to be the badly estimated $\gamma O \pi^0$
vertex. It seems therefore that the $f_2$ production process would be based on more solid
estimates of the coupling.

\section{Diffractive $\eta_c$ photo and electro-production}
In order to apply, to some extend, perturbative QCD to the calculations, one can look
at processes where heavy quarks are involved.
Diffractive $\eta_c$ production in DIS has been studied at lowest order
\cite{Czyzewski:1997bv,Engel:1998cg}.
The calculations give $\sigma \approx 11$ pb at $Q^2=0$ and $0.1$ pb at
$Q^2=25$ GeV$^2$, with no energy dependence.

This process has been recently reanalyzed \cite{Bartels:2001hw} in LLA.
The amplitude (\ref{ampli}) has been calculated in the saddle point approximation
using the Green function $G_3$, constructed with the non forward Odderon states in
(\ref{oddwave}). One has
\bea
G_3(y)=
\sum_{{\rm odd}\ n}\int_{-\infty}^{+\infty} d\nu e^{y\, \chi(\nu,n)}
N(\nu,n) E_3^{(\nu,n)} {E_3^{(\nu,n)}}^* \, ,
\label{greenf}
\eea
where $N(\nu,n)$ fix the representation dependence of the measure\cite{Bartels:2001hw}.
In order to compare the effect of LLA QCD resummation to the lowest order calculations,
the same impact factors for the $\gamma O \eta_c$ and for $p O p$ vertices have been used.
The $\gamma O \eta_c$ impact factor has been computed perturbatively
\cite{Czyzewski:1997bv}; it has an interesting symmetry which allows a partial
analytical computation\cite{Bartels:2000yt,Bartels:2001hw} of its scalar product with the
Odderon eigenstates in (\ref{oddwave}).
For the proton side the ansatz previously made\cite{Czyzewski:1997bv} has been used.

\begin{figure}[t]
\epsfxsize=12pc 
\centering
\begin{picture}(150,50)
\put(0,0){\epsfbox{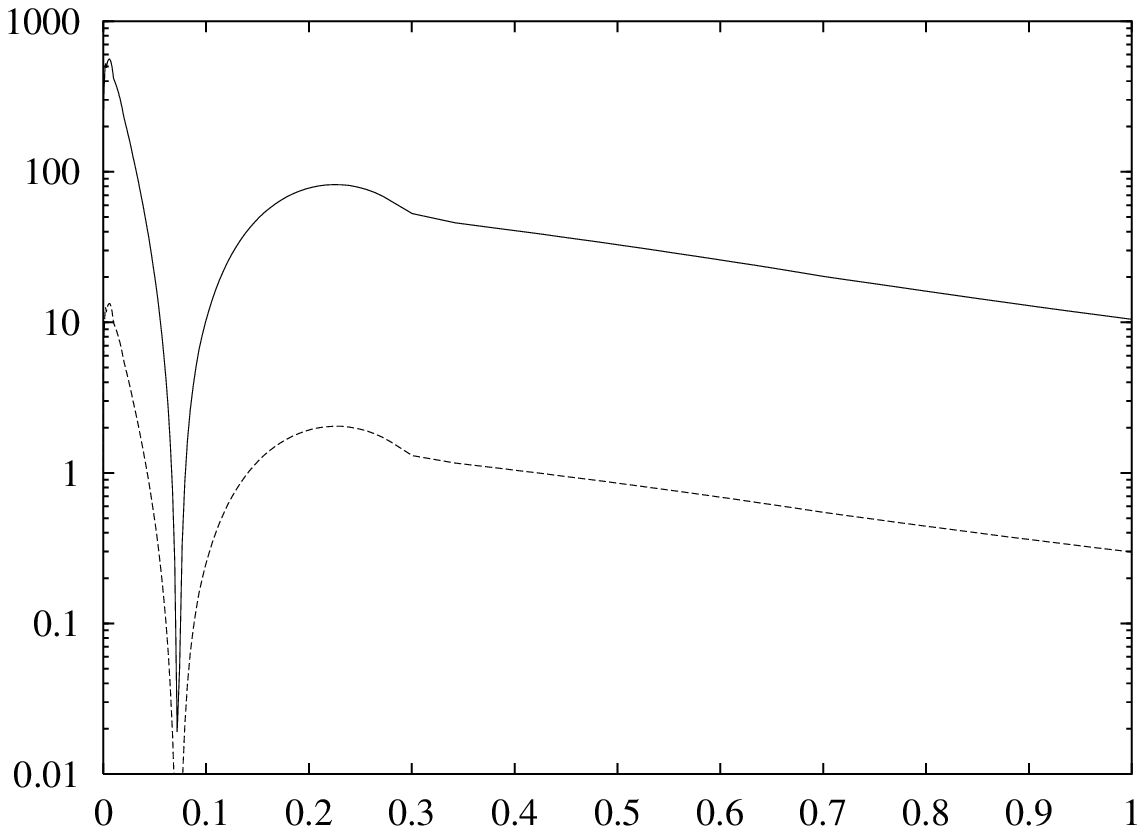}} 
\put(-5,50){$ \displaystyle{d\sigma \over dt}$ }
\put(140,0){$|t|$}
\end{picture}
\caption{The differential cross sections (in pb $/$ GeV$^2$). 
The upper curve refers to $Q^2=0$.
\label{Fig1}}
\end{figure}

Due to the structure of the Odderon states, which manifests as
a strong correlation between the constituent reggeized gluons, the Odderon
coupling to the impact factor has the dominant real part which changes sign for a
value of the momentum transfer squared $t$. The computation of the differential cross section
leads therefore to the result presented in Fig. \ref{Fig1}, where a dip in the small $t$ region
is present. Due to the cut nature of these Odderon singularities the cross section
is slightly suppressed (as $1/\ln{s}$) with energy.

The total cross section, which results from the LLA Odderon states contribution, has been found
to be  $\sigma \approx 50$ pb at $Q^2=0$ and $1.3$ pb at $Q^2=25$ GeV$^2$,
an order of magnitude larger than in the simple three gluon exchange case. 
Quantitative and qualitative remarkable differences are
introduced by the gluon interaction but the cross sections are small to be measured at HERA.

The Odderon still represents a challenge for theory and experiments.

\section*{Acknowledgments}
The author is very grateful to J. Bartels and H.G. Dosch
for very interesting and helpful discussions.

\end{document}